\documentclass[bibyear]{aa}
\usepackage[varg]{txfonts}
\usepackage{natbib,twoopt}
\usepackage[breaklinks=true]{hyperref} 
\bibpunct{(}{)}{;}{a}{}{,} 
\begin{document}

\title{Dust cleansing of star-forming gas: II. Did late accretion flows change the chemical composition of the solar atmosphere?} 

\author{Bengt Gustafsson \inst{1, 2}}

\offprints{Bengt Gustafsson, \email{bengt.gustafsson@physics.uu.se}}

\institute{Department of Physics and Astronomy, Uppsala University, Sweden
  \and Nordic Institute for Theoretical Physics (NORDITA), Stockholm, Sweden} 

\date{Received 03/05/2018 / Accepted 31/08/2018}

\abstract {}{The possibility that the chemical composition of the solar atmosphere has been affected by radiative dust cleansing of late and weak accretion flows
by the proto-sun itself is explored.} {Estimates,
using semi-analytical methods and numerical simulations of the motion of dust grains in a collapsing non-magnetic and non-rotating gas sphere with a central light source
are made in order to model possible dust-cleansing effects.} {Our calculations indicate that the amount of cleansed material may well be consistent with
the abundance differences observed for the Sun when compared with solar-like stars and with the relations found between these differences and
the condensation temperature of the element. } {It seems quite 
possible that the proposed mechanism produced the significant abundance effects observed for the Sun, provided that late and relatively weak 
accretion did occur. The effects of cleansing may, however, be affected by
outflows from the Sun, the existence and dynamics of magnetic fields and of the accretion disk, and the possible presence and location of the early Sun in a rich stellar cluster.}  

\titlerunning{short}
\maketitle
%
\section{Introduction}

The results by \citet{Melendez09}, showing that the Sun departs from most solar twins in the solar neighbourhood in that the composition of its surface layers is comparatively
rich in volatile elements and poor in refractory elements, have led to considerable discussion (for a recent review see \citet{Adibekyan17}). Although the finding, a correlation
between the chemical abundances for the Sun
relative to the twins with the condensation temperatures of the elements in a gas, has been verified by several independent studies \citep{Ramirez09, Gonzalez10, Adibekyan14, Nissen15, Spina16}, its interpretation is
still disputed. The early suggestion by \citet{Melendez09} that it could be due to the formation of terrestrial planets in the proto-planetary nebula before the latter
was dumped onto the Sun, polluting its convection zone by dust-cleansed gas, requires either an unusually long-lived nebula or a very rapid retraction of the
solar convection zone to the surface (cf \citet{Gustafsson10}). As suggested by \citet{Onehag11} this early retraction might result from the episodic accretion scenario according to
\citet{Baraffe10}, if applied to the Sun but not to the twins. The alternative, that is, that the effect is due to Galactic chemical evolution, and migration of the Sun from other
galactic regions than those of the twins, was suggested and partly supported by tendencies found by \citet{Adibekyan14}, \citet{Maldonado15}, \citet{Maldonado16}, and \citet{Nissen15} 
for some correlation of the abundance effects to depend on stellar ages. Still another possibility was offered by \citet{Gustafsson10} who proposed that the pre-solar nebula was 
cleansed early on from dust by the radiation pressure from hot stars in the neighbourhood. This was supported by the results of \citet{Onehag11} and \citet{Onehag14} who carried out detailed analyses of solar-twins and solar analogues in the rich, old cluster M\,67, and found the cluster stars to have an abundance profile more similar to that of the Sun than most solar twins in the neighbourhood. However, \citet{ Gustafsson18} recently found that the cleansing mechanism proposed  was unlikely to be efficient enough to explain the chemical peculiarities of a relatively massive cluster like M\,67. Also, the recent finding of binary stars with one component showing
deviating abundances from that of the other, such as 16 Cyg \citep{Tucci14, Nissen17}, XO-2N/XO-2S \citep{Ramirez15, Biazzo15}, WASP 94 A/B \citep{Teske16a}, HD\,133131 A/B \citep{Teske16b}, HAT-P-4 \citep{Saffe17} and the co-moving pair HD240430/HD1240429 \citep{Oh17}, and sometimes
 with differences correlating with the condensation temperature of the element, suggests that the phenomenon may be local. 

An interesting fact, first noted by \citet{Gonzalez10}, is that the meteoritic elemental abundances show a similar tendency with the condensation temperature to the solar photospheric abundances, in the sense that the meteorites tend to have an even lower ratio of refractory elements relative to volatiles, than the 
photosphere. This also suggests that the phenomenon should at least partially be local. 

The issue studied in the present paper is whether radiative cleansing of the matter accreted onto the Sun could be provided by the 
proto-Sun itself, in its accretion flow. For this mechanism to work, the friction forces between the dust-grains and the gas as well as the gravity forces must be balanced by 
the radiative forces. Since the friction is dependent on the flow, this mechanism requires constraints on the density of the flow and on its infall velocity in order to work. 
Other constraints will be important, such as those  on the luminosity of the proto-Sun and on the extinction in the cloud around it. We shall use elementary considerations to explore 
these constraints and shall then apply a somewhat more detailed model to quantify the possibilities of the suggested mechanism to explain the observed abundances in the presence of turbulence in
the cloud. 

Our model mimics an initially homogeneous
cloud of neutral atomic or molecular gas with dust, and with a light source representing the early Sun in its centre. The aim will be to find out of whether or not the
radiative pressure from the Sun is able to prevent the dust, and thus the chemical elements in the dust, to fall onto the star. 


\section{The model and the results}

We consider a specific dust grain, assumed to be spherical with radius $a$ and mass $m$. We set the vector, extending in space from the Sun to the 
grain, ${\bf r}$, with a unit vector ${\bf e_r}$ in the direction of ${\bf r}$. In interstellar clouds, the mean free path of the molecules is orders of magnitude longer than the size of the dust grains.
In this case the equation of motion for a grain with mass $m$, illuminated by the Sun with luminosity $L$ and mass $M$, can be written (see \citet{Draine11})
\begin{equation}
\begin{split}
m \cdot \frac{{\rm d}^2 {\bf r{\rm (t)}}}{{\rm d}t^2} = \left[\frac{Le^{-\tau}\pi\,a^2\,\overline{q_{pr}}}{4\pi  c} - G\,m\,M\right]\frac{\bf e_r}{r(t)^2}-  \\
- 2\pi a^2 \,n(t)kT\, {\bf e}_{\bf v,v_g}B(s)  
\label{eq0}
\end{split}
.\end{equation}
The three terms on the right-hand side (RHS) represent the radiative force, the gravity, and the friction force exerted by the gas on the grain, respectively.
The electromagnetic radiation is attenuated by extinction, and the grain is located at an optical depth $\tau$ from the Sun at time $t$. The factor 
$q_{pr}$ is numerical, and is dependent on the grain composition and shape, and on wavelength. For the relevant
dust and stellar light discussed here a mean $\overline{q_{pr}}\equiv q$ of about 1.5 
may be adopted, see, however, Fig 1 of \citet{Ferrara91}, indicating that a higher value up to $q=2.5$ could be appropriate for
minor graphite grains. Similarly, if the grains have a composite fluffy character, as suggested by \citet{Kimura17} on the basis of captured Local Interstellar Cloud grains, an increase of $q$  by up to a factor of two may be adequate
for the more massive particles.
The speed of light is $c$ and Newton's constant of gravity is $G$.  
The dust grain is retarded by a drag force, described by the last term of the right-hand side of Equation (\ref{eq0}). 
The number density of atoms or of molecules is $n(t)$, the kinetic temperature
is $T,$ and $k$ is Boltzmann's constant. Following \citet{Draine11} we may write
\begin{equation}
B(s)\approx \frac{s}{\alpha}[1+(\alpha\cdot s)^2]^{1/2} + \left(\frac{eU}{kT}\right)^2 {\rm ln}\Lambda \frac{s}{3\sqrt{\pi}/4 + s^3},
\label{eq101}
\end{equation}
\begin{equation}
s \equiv \frac{v_{\rm drift}}{\sqrt{2kT/m_{\rm H}}},  \,\,\,\alpha \equiv 3\sqrt{\pi}/8, \,\,\, \Lambda = \frac{3kT}{2a e |eU|}\left(\frac{kT}{\pi n_e}\right)^{1/2}, 
\label{eq102}
\end{equation}
where $v_{\rm drift} {\bf e_{v,v_g}}={\bf v}-{\bf v_{\rm g}}$ is the drift velocity of the grain relative to the gas in the direction ${\bf e_{\rm v,v_g}} $, 
with   ${\bf v} = {\rm d}{\bf r}/{\rm
d}t$ and ${\bf v}_g$ being the velocities of the grain and the gas, 
respectively, and  $m_{\rm H}$  the mass of the hydrogen atoms or molecules, whatever species is assumed to dominate the gas. 
In the expression for $B(s)$ in Equation (\ref{eq101}), the last term represents the Coulomb forces, 
where $U$ is the grain potential, $e$ the electron charge, and $n_e$ the
electron density in the gas. The potential is determined by the electric charge on the grain. Neglecting photoelectric emission, this charge is set by the
requirement that the positive ions captured per time unit be balanced by the number of electron captured. For small degrees of ionization one then finds 
from Equations (9-15) of \citet{Spitzer78} that $eU/kT=-3.76$. One may schematically estimate the photoelectic emission (positive) contribution to the potential, as a result of the interstellar ultraviolet (UV) background.
Following  \citet{Spitzer78}, p. 144-146 and his Equation (6-20), setting the sticking probability of electrons striking the grain 
and the absorption efficiency factor equal to 1, and adopting a value of  0.1 for the photoelectric efficiency of radiation with wavelengths in the interval 91-110 nm, 
and else much smaller, and assuming the optical depth to be small, we obtain a photoelectric contribution to  $eU/kT$ of:
\begin{equation}
\left(\frac{eU}{kT}\right)_{ph} \approx 2\cdot 10^6\left(\frac{\pi m_e}{8 kT}\right)^{1/2}\frac{1}{n_e} - 1.
\label{eqnS1}
\end{equation}
To this we should add a contribution due to the radiation from the early Sun. We take the flux in the relevant UV interval of the present Sun from \citet{Lean87} to be 
representative, and then multiply the right-hand side of Equation (\ref{eqnS1}) by a factor $(1+2.2\cdot 10^{30}/r^2)$ where the distance to the Sun is $r$ in centimetres. 
For our neutral gas  (with degrees of ionization $n(e)/n(H) =2.72 \cdot 10^{-4}$ at $T=160$\,K  from Table 30.1 in \citet{Draine11b}) we find the Coulomb force term to be relatively small for the local contribution from the 
electron-ion balance. However, we shall see that the photoelectric charging of the grains has some significance. 

We note that the dependence of the drag force on the drift velocity is linear for small speeds -- in typical cases with $T = 10$\,K the shift 
to a mainly quadratic dependence occurs around $v_{\rm drift}\sim 400$ m/s which is higher than typical sound speeds in the cool neutral gas but smaller
than typical free-fall velocities for gas at distances of several thousand astronomical units (AU) from the Sun. 

We assume that the gas velocity is not affected notably by the friction of the grains, and is set by 
\begin{equation}
{\bf v_{\rm g}} = - \sqrt{\frac{2 G M}{r}}\times {\bf e_r} + {\bf v_{\rm turb}}
\label{eq103}
,\end{equation}
where the free-fall velocity in the first term is corrected by the fluctuating turbulent velocity of the gas, ${\bf v_{\rm turb}} = {\bf v_{\rm turb}}({\bf r},t)$. For the gas density we assume spherical symmetry
around the Sun and a time-independent situation, that is, 
\begin{equation}
n({\bf r},t) = n(r) = n_0 \cdot \left(\frac {r_0}{r   }\right)^\beta
\label{eq104}
,\end{equation}
where $\beta$ is usually chosen to be $3/2$, corresponding to a stationary flow, or alternatively $2$, as suggested by \citet{Shu77}, and with the density profile of an isothermal
sphere. There is observational support for such power-law density profiles for collapsing proto-stellar clouds and low-mass protostars (e.g. \citet{Loren83}, \citet{Jorgensen02}, \citet{Persson16}), but they are singular at $r=0$ and invalid close to the star. We have stopped the integration at $r=100$ AU to take the existence of
an accretion disk into account. In solving
Equation (\ref{eq0}), we alternatively set the initial velocity of the grain to ${\bf v}_{\rm 0} = {\bf 0}$ or ${\bf v}_{\rm 0} = - \sqrt{2GM/r_0}\cdot {\bf e_r}$, that is, the free-fall velocity, at $t = 0$ and $r = r_0$.  The factor $r_0$ is set to 
1000 AU and alternatively 10,000 AU. The number density of hydrogen atoms/molecules at $r_0$ is calculated to correspond to a given accretion rate $\dot{M}$ at $r_0$, calculated with the assumption of the free-fall speed from zero speed at infinity and a stationary flow which leads to
\begin{equation}
n_0 =  \frac{\dot{M}}{4\pi \sqrt{2GM}r_0^{3/2}\mu m_{\rm H}},
\label{eq105}
\end{equation}
where $\mu$ is $1+(Y+Z)/X$ with $X$, $Y$ and $Z$ being the mass fractions of hydrogen, helium, and elements heavier than helium, respectively, in the gas.
We end the calculation when the dust grain has reached a certain inner radius $r_1$; here, the neglect of a net angular momentum for the system is certainly 
inadequate, and the dust should in reality have ended up in an accretion disk. We also stop the integration at a time of 6 Myr. Then, in practice dust grains that have not reached the inner radius are pushed out to relatively large distances from the Sun by the radiative forces and are regarded as lost from the system.

\subsection{The non-turbulent case}
We shall first assume the turbulence velocities to be zero, that is, ${\bf v_{\rm turb}}$ is set equal to ${\bf 0}$ in Equation (\ref{eq103}). In this case the flows of grains and gas will
constantly be directed towards the Sun, and the model will remain spherically symmetric. The equation of motion is simplified to 
\begin{equation}
 \frac{{\rm d^2} r}{\rm dt^2} =\left[\frac{3}{16 \pi} \frac{ Le^{-\tau}\,q}{c a \rho_d }-G\,M\right] \cdot \frac{1}{ r^2 }-  \frac{3}{2} \frac{ \,n(t)kT}{a \rho_d }\cdot B(s) 
\label{eq1}
,\end{equation}
where $\rho_d$ is the mass density in the dust grains. 

The present discussion is focussed on the conditions in the gas at a considerable distance from the Sun during the last accretion in its formation history, assumed to entail relatively thin gas. Therefore, an approximation made in our estimates below is the assumption of small optical depths; for example, we assume $\tau=0$ in Equations (\ref{eq0})
and (\ref{eq1}). In a subsection below, 
we explore the adequacy of this assumption. 

Two necessary conditions for the dust to be accelerated outwards are seen directly from Equation (\ref{eq1}). The radiative force has to be greater than the 
gravity force on the grain, that is, the large parenthesis on the RHS has to be positive, such that
\begin{equation}
a \leq \frac{3\,L \,q}{16 \pi \,\rho_d\,c\, G\,M} .
\label{eq2}
\end{equation}
With characteristic numbers for the parameters ($L = 0.5 L_\odot$, $q=1.5$, $\rho_d=1$ g/cm$^3$ , and $M=1 M_\odot$) we find $a_{max} = 4.5\cdot 10^{-5}$ cm.
This constraint implies that, with a dust size distribution $N(a) \sim a^{-3.5}$, as is often assumed following \citet{Mathis77}, see also \citet{Casuso10}, and extending from $a=10^{-7}$ to $10^{-4}$ cm$^{-3}$, more than half of the total dust mass in the accreting flows may be pushed out by radiation with this choice of parameters. 

The second necessary condition for an outward acceleration is that the radiative force in Equation (\ref{eq1}) is stronger than the friction force due to the down-streaming gas. Since $|v_{\rm drift}|>\sqrt{2GM/r}$ for outgoing grains and neglecting the electric forces, we find from Equations (\ref{eq101}), ({\ref{eq102}), and (\ref{eq1})
\begin{equation}
\dot{M} \leq \frac{q\mu}{c\sqrt{2G}}\, \cdot L \left( \frac{r}{M}\right)^{1/2}; \,\,\, v_{\rm drift} \gg \sqrt{2kT/m_{\rm H}}.
\label{eq33}
\end{equation}
With characteristic numbers as above, and with $T=10$K, we find from Equation (\ref{eq33}) a maximum accretion rate of  0.050 M$_\odot$/Myr  
at $r=10^4$ AU or 0.016 M$_\odot$/Myr at $r=10^3$ AU. For small graphite grains with $q \sim 2.5,$ 
 maximum accretion rates that are almost double this latter rate may be possible. If we require a significant contribution to the star by dust-rarefied gas with a mass of at least the mass of 
the present convection zone (or more, since the  convection zone during the first millions of years of the Sun should have been significantly more massive) we have to assume that the last percents of the solar mass 
were accreted slowly, during a time of several hundred thousand years or more in order to satisfy the condition of Equation (\ref{eq33}). We note that such accretion rates are at least two orders of magnitude smaller than those
of the zoom-in simulations of forming accretion disks in giant molecular clouds (GMCs) by \citet{Kuffmeier17}, and that the timescale of our slow accretion extends far beyond that of these authors
as well as the timescale of \citet{Shu77}. 
 
The relative significance of the radiative pressure and the gravity is not dependent on the distance from the star if the optical depth is small, as is demonstrated by 
the terms within the parentheses in Equation (\ref{eq1}). The $r^{1/2}$ dependence in Equation (\ref{eq33}) implies, however, that the upper limit on $\dot{M}$, set by the requirement that dust grains should be stopped from falling inwards by
radiation pressure, becomes tighter  as the dust grain
is brought closer to the star (e.g. by initial momentum or by turbulence).  This means that if moving inwards in
a given free-fall flow of gas, the dust grains may finally come to a radius $r_c$ where they cannot escape from falling into the accretion disk. 
For the case
with high drift speed, we obtain
\begin{equation}
r_c =\frac{1}{8\mu^2} \dot{M}^2 \cdot GM \cdot \left[\frac{Lq}{4c}-\frac{4\pi}{3} G M a\rho_d\right]^{-2}.
\label{eq34}
\end{equation}
With characteristic parameters, as above, and with $\dot{M} = 0.01$ M$_\odot$/Myr and $a = 2\cdot 10^{-5}$ cm, we find $r_c=1.5\cdot 10^3\,$AU. In practice this means that the
dust that separates from the infalling gas will have to decouple from the gas already at distances on the order of 1000  AU from the proto-Sun or further out. It is 
worth noting that these distances overlap with the location of the Oort cometary cloud. Also, the last factor of the last term in Equation (\ref{eq34}) suggests a 
possible fractionation of dust of different sizes: with the characteristic parameters chosen above, $r_c$ 
= 480\,AU for grains with $a= 0.35\cdot 10^{-5}$ cm, while 
$r_c$  = 13,000 AU for $a=3.5\cdot 10^{-5}$ cm. 

We have explored the possibilities  for dust-grains to exist at various distances from the proto-Sun by setting the LHS equal to zero in Equation (\ref{eq1}) and then solving
it for different values of $\dot{M}$, $a$ and $r$. The results for the non-electric case are displayed as full lines in Figure \ref{self_fig1}, where the regions in the diagram below each curve are the `allowed' regions, that is, where dust grains will be pushed out by radiation and not pulled by gravity and friction from the in-falling gas towards the stellar accretion disk. The corresponding curves for the
case when electric forces are included (dashed blue lines in Fig. \ref{self_fig1}) clearly show that the critical gas accretion rates, able to sweep dust grains close to the star, are reduced
by almost a factor of two, as compared with the case when coulomb forces are neglected. In particular, the results of considering the photo-electric effects on the grains are noteworthy. The effects are particularly important for the larger grains which contribute significantly to the totally accreted mass of dust. 

\begin{figure}
   \resizebox{\hsize}{!}{\includegraphics{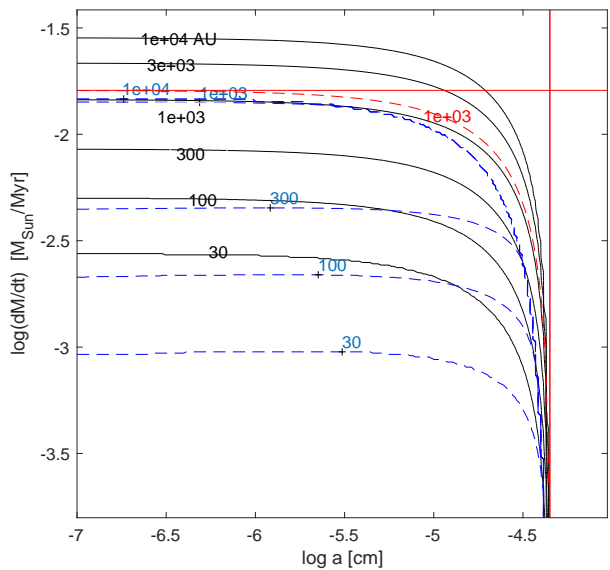}}
  \caption{The "allowed" parameter space (under each curve) for dust grains with radius $a$ that may exist around a star at a certain distance (indicated in AU at each curve) without being accreted to the star by gravity and friction from the in-falling gas with different accretion rates $\dot{M}$. The full lines correspond to cases with neutral grains while dashed blue curves indicate
corresponding cases with electrically charged grains. The approximate upper limit on $a$, according to Equation (\ref{eq2}), is indicated by a vertical red line, while the
  approximate upper limit on ${\rm d}M/{\rm d}t \equiv \dot{M}$ for a distance from the star of $10^3$ AU, according to Equation (\ref{eq33}), is indicated by a horizontal red line. Also shown is 
  the approximate critical radius $r_c$ as calculated by Equation (\ref{eq34}) as a dashed red curve. The fair agreement between the approximate (in red) and the more exact (black) curves is due to the fact that the quadratic term in the dependence of the friction force on drift speed, c.f. Equation (\ref{eq101}), is indeed dominating, as assumed in the approximations. The reduction of
  the "allowed" accretion rates for the charged grains relative to the corresponding cases with neutral ones is obvious.}  
  \label{self_fig1}
\end{figure}

\subsection{The optical depth effects and the friction forces onto the gas}

Above, we have assumed that the medium is transparent to the radiation from the star in the centre, thus assuming that the dust grains 
close to the star do not shadow the grains behind. In order to take such possible shadowing effects into consideration, we shall calculate the optical depth in the
stellar envelope. We then disregard the effect of the dust in the accretion disk, which may well be optically thick in certain directions. The optical
depth $\tau$ in the converging accretion flow, assumed to be spherically symmetric around the star, is
 \begin{equation}
 \tau = \int _{r_1} ^r {\rm d}r \int _{a_l }^{a_u} N_{\rm V}(a,r,t) \sigma(a) {\rm d}a ,  
  \label{eq1531}
 \end{equation}
where $\sigma(a)$, which we here set equal to $\pi a^2$, is the absorption cross-section per grain, and 
$N_{\rm V}(a,r,t)$ is the number of dust grains with radii in the interval $(a,a+{\rm d}a)$ per volume unit at time $t$. When calculating $N_{\rm V}$ we have solved the 
equation of motion (Equation (\ref{eq1})) for a range of different $a$ values from $a_l=10^{-7}$ cm to $a_u = 10^{-4}$ cm, distributed in size in proportion to $\sim a^{-3.5}$ according to \citet{Mathis77}, in a convergent 
and stationary gas flow in free fall (i.e. applying Equation (\ref{eq103}) with ${\bf v_{\rm turb}=0}$ and with $\beta=3/2$ in Equation (\ref{eq104})). The gas-to-dust mass 
ratio was assumed to be 100, however, noting that there may be considerable variations in this ratio in interstellar clouds (see, e.g. \citet{Reach15}).  As initial
conditions we have alternatively chosen ${\bf v(r_{\rm 0}) = 0}$ and ${\bf v(r_{\rm 0}) = v_{\rm ff}(r_{\rm 0})}$, with ${\bf v_{\rm ff}}$ denoting the free-fall velocity with a start at rest in infinity, however with little
effects on the end results. Typical values of $r_{\rm 0}$ and $r_1$ were 10,000 and 1000 AU, respectively.
The individual grain orbits have been integrated for at least 1 Myr each in order to
ascertain that a stationary situation has been achieved. 

When calculating $\tau$ we have used an iterative procedure, first setting $\tau(r) = 0$ when solving Equation (\ref{eq1}), subsequently calculating $\tau(r)$ by 
Equation (\ref{eq1531}), feeding the result into Equation (\ref{eq1}), and so on. In these calculations, we have neglected the coulomb forces in $B(s$).
In practice, with the small resulting $\tau$ values, we have found that three
iterations are enough to ascertain a satisfactory convergence.The resulting optical depths for three different accretion rates are shown in Fig. \ref{self_fig2}. As is seen, 
at these rates the depths are relatively small or even negligible. 

\begin{figure}
  \resizebox{\hsize}{!}{\includegraphics{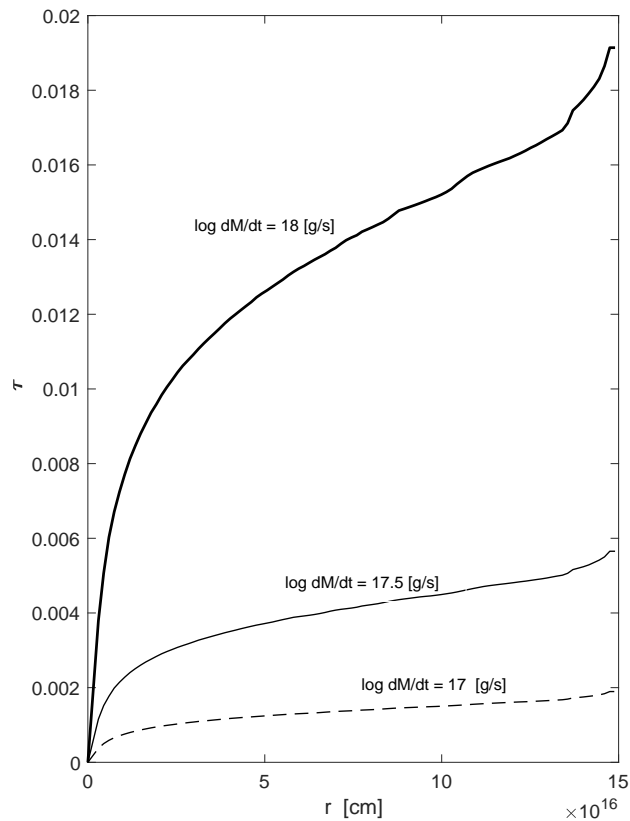}}
  \caption{ The optical depth as a function of distance from the star for three different accretion rates, $10^{17}$ g/s, 3 x $10^{17}$ g/s and $10^{18}$ g/s, corresponding to 0.0016, 0.0047 and 0.016 M$_\odot$/Myr, respectively.}
  \label{self_fig2}
\end{figure}

\begin{figure}
  \resizebox{\hsize}{!}{\includegraphics{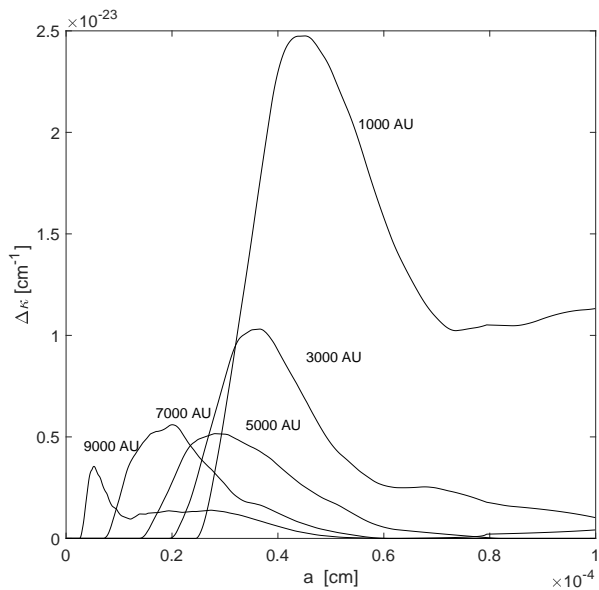}}
  \caption{The contributions to the dust opacity for grains of different radii $a$, at different distances from the accreting star indicated at each curve in AU. The opacity
  contribution is given in cm$^{-1}$, per logarithmic interval in $a$. The dominating contribution from grains on the order of $a=a_{\rm max} $ in size, cf. Equation (\ref{eq2}),
  is seen in the inner regions, and is enhanced by the convergence of the flow, while at greater distances from the star the more numerous smaller grains contribute more, as      
  a result of their initial motion inwards due to the starting condition $v(r_{\rm out}) = v_{\rm ff}$. The curves have been smoothed by a 
  filter with a width of $5 \times10^{-6}$ cm, to compensate for fluctuations due to the stochastic nature of the method used. }
  \label{self_fig3}
\end{figure}

A basic reason for the small optical depths is the fact that the numerous grains with small radii are already pushed out by the stellar radiation at great
distances from the star, where the gas flow has still not become strong and dense enough to pull them inwards by friction. The dominating dust opacity is instead
contributed by the more massive grains, in particular those that have $a$ close to $a_{\rm max}$ as estimated in Equation (\ref{eq2}). The close balance
between radiative forces and gravity makes
them fall slower towards the star than the more heavy ones, which they also outnumber due to the $a^{-3.5}$ dependence of the initial number density. 
The situation is illustrated in Figure \ref{self_fig3}, where the opacity contribution by grains of different sizes and at different distances from the star 
is illustrated. 

In addition to neglecting dust opacities in the model calculations, we have neglected the friction forces from the dust onto the gas. Applying Newton’s third law, 
one easily finds in the high drift-velocity case that
the friction force per gas mass, relative to the corresponding gravity force, is
\begin{equation}
\frac{F_{\rm fric}}{F_{\rm grav}} = \frac{3}{64\sqrt 2\pi} \cdot \frac{{\dot M} v_{\rm drift}^2}{(GM)^{3/2}(a_ua_l)^{1/2} \phi \rho_d \mu} r^{1/2},
\label{eq881}
\end{equation}
where $\phi$ is the gas-to-dust mass ratio. 
By setting in representative numbers one finds that for our models this ratio is on the order of $10^{-2}$ or less, which means that the friction forces
on the gas can be neglected. It is obvious, however, that for considerably more intensive gas flows, such that ${\dot M} > 0.1 \,M_\odot$ per million years, or high 
luminosities increasing the radiative forces and thus $v_{\rm drift}$, the friction forces might become significant.    

\subsection{Conclusions from the spherically symmetric model}

It is clear from the results presented above that a slow accretion of gas, with an accretion rate on the order of a few percent of a solar mass per million years, 
should enable the central star to push most of the grains with radii less than about $5\times 10^{-5}$ cm, corresponding to roughly half the grain mass,
out of the accretion gas flow.  A considerably 
higher accretion rate will both increase the friction forces and the optical depth of the envelope, 
due to relatively large grains, so much that the smaller grains fall inwards, and
the cleansing becomes significantly weaker. Thus, a model with a moderate accretion flow
seems viable to explain stellar atmospheres depleted in dust-forming elements for stars with convection zones of a few percent of the stellar mass,
provided that such relatively thin convection zones exist at times when this mild accretion is still active. According to standard models of PMS
evolution (\citet{Montalban08}) the convection zone remains massive, with masses greater than about 0.1 M$_\odot$ until about 20 Myr, while typical accretion gas disks only remain for less than 10 Myr (see, however, \citet{Pfalzner14}). Thus, we may have a timescale problem similar to that discussed for the planet-cleansing hypothesis 
for the abnormal volatiles/refractories ratio of the Sun  \citep{Melendez09, Gustafsson10}. 
This could possibly be explained as a result of episodic accretion or more long-lived stages of the late accretion disks. 

Another possible caveat of the present model is, however, turbulence which was found by \citet{Gustafsson18} to be a major problem with the alternative hypothesis of massive dust-cleansing by bright stars 
as an explanation for the Mel{\'e}ndez effect. The effects of turbulence on the present self-cleansing model are discussed below. 

\subsection{The turbulent case}

\begin{figure}
  \resizebox{\hsize}{!}{\includegraphics{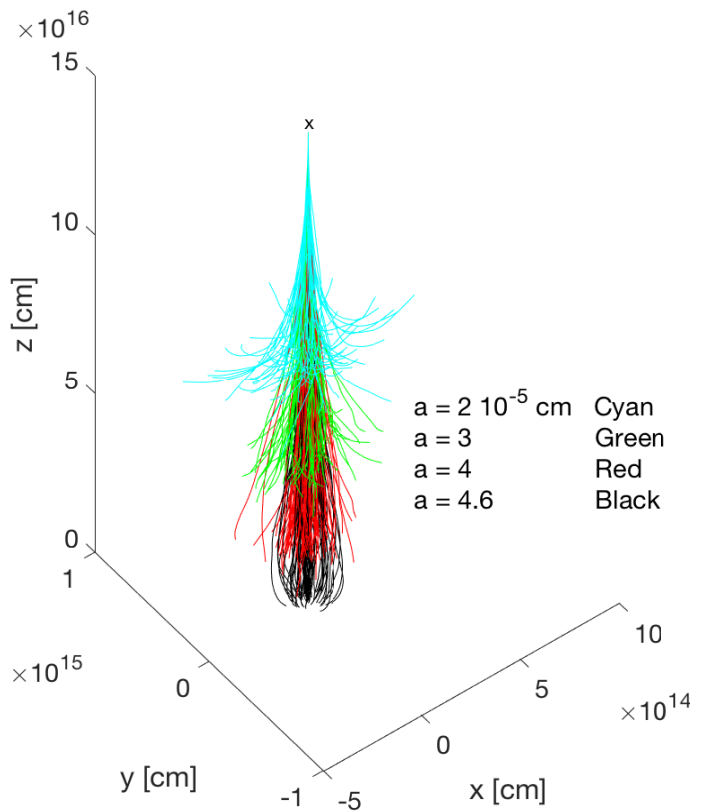}}
  \caption{The orbits of grains over $10^5$ yr, from a point indicated by an `x' in black along the z axis at a distance of $10^4$ AU from the star (which is 
at the origin of the coordinate system), and with different grain radii $a$ as indicated by the different colours. The dust grains were given an initial velocity
identical to the gas free-fall velocity. The turbulent mean speed was set to 0.1 km/s. We note that the scales of the x and y axes are different from that of
the z axis.  }
  \label{self_fig3a}
\end{figure}
\begin{figure}
  \resizebox{\hsize}{!}{\includegraphics{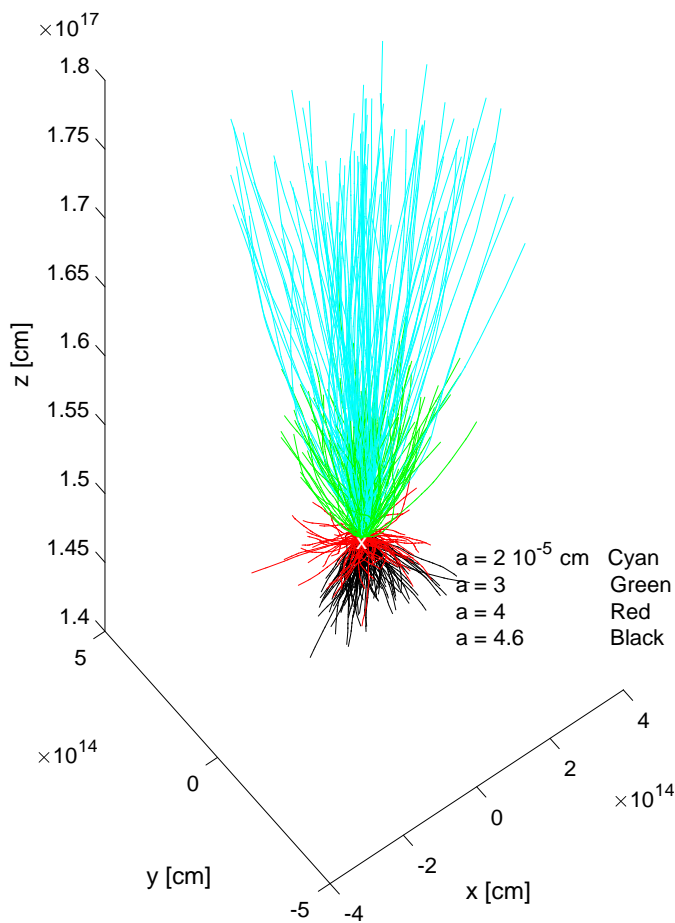}}
  \caption{The orbits of grains over $10^5$ yr, from a point indicated by an `x' in white along the z axis at a distance of $10^4$ AU from the star (which is 
at the origin of the coordinate system), and with different grain radii $a$ as indicated by the different colours. The initial velocity of the dust grains was set
to zero. The turbulent mean speed was set to 0.1 km/s. We note that the scales of the x and y axes are different from that of
the z axis. }
  \label{self_fig4}
\end{figure}

The origins and the consequences of the motions in giant molecular clouds are still discussed intensively. Observationally, however,
it seems clear that turbulence is present in star-forming regions as well as quiet clouds.
\citet{Solomon87} showed that the CO line widths of the GMCs scaled with the square root of their radii $R$, according to $\sigma_{\rm v} \approx 0.72 (R/{\rm pc})^{0.5}$ km/s, which seems to hold down to small scales (cf \citet{Bolatto08}, \citet{Rosolowsky08}, \citet{Heyer04}). The ideal value of $\sigma_{\rm v}$ to be chosen for a model
of the solar birth cloud is therefore, however, highly uncertain. We therefor treat it as a free parameter here, varying it between 0.1  and 5 km/s.
To handle the turbulence in the accretion flow of the gas in our model in full detail is beyond the scope of the present paper. Instead, we have followed
the schematic recipe of \citet{Higham01}, by re-setting the turbulent velocity, ${\bf v}_{\rm turb}$, when
solving Equation (\ref{eq0}) numerically every time the derivatives are to be calculated for this solution. We have used the MATLAB routine ODE23 for this purpose, which is based on a Runge-Kutta method with variable
(and continuously adjusted) step size in time. The step size is typically 30 yr, which is less than one thousandth of the free-fall time from 10,000 AU. We thus set   
\begin{equation}
{\bf v}_{\rm turb} = \xi_{\rm x}{\bf e}_{\rm x} + \xi_{\rm y}{\bf e}_{\rm y} + \xi_{\rm z}{\bf e}_{\rm z}
\label{eq1541}
\end{equation}
in Equation (\ref{eq103}), where $\xi_{\rm x,y,z}$ are three independent random numbers drawn for each time step from a Gaussian distribution with a dispersion $\sigma_{\rm v}$. 

Another effect of turbulence is a fluctuating gas density. In order to study the effects of such fluctuations, we have run some models with 
the density perturbations guided by the log-normal distribution found from detailed simulations by \citet{Padoan02} (their Equation 6), around values calculated according to Equation (\ref{eq104}), with a characteristic fluctuation time of  $10^4$ years.

At the solution of Equation ({\ref{eq0}) we have set the optical depth equal to zero, guided by our results in
Section 2.2. Subsequently, the 
solution entails solving three coupled second-order non-linear differential equations. For each run, this has been 
done for 1000 different dust grains of given radii $a$, initially located at a distance of $r_0=10,000$ AU from the star (along the positive z axis). 
In subsequent runs, $a$, $\sigma_{\rm v}$, ${\dot M}$, and the initial condition have been varied, one at a time. For the initial velocity of the dust grain
at the starting distance we have alternatively used ${\bf v(r_{\rm 0})} = {\bf 0}$ and ${\bf v(r_{\rm 0})}=-\sqrt{2GM/r_0}\cdot {\bf e}_{\rm r}$. The integration proceeds for 5 Myr. If the orbit during this time gets closer to the star than 100 AU, we assume the grain to be caught in the 
accretion disk around the star, and finally fall into the star. Conversely, if it does not get closer to the star during this time we consider it lost. Grains that belong to this class are usually accelerated to very large distances from the star by the radiative pressure. One should note that in these simulations we totally neglect the outflows
from the central disk, known observationally to carry away a considerable fraction of the mass; see \citet{Alexander14} and references therein. 

In most of these simulations for the turbulent case we have neglected the effects of coulomb forces on the grains. To check their significance, however,  we made a series of simulations, 
taking the charging of the grains into account. With the parameters of these effects set as described above, they are found to 
have relatively limited effects on the results, the most important one being that due to the photoelectric charging. This is further illustrated below.

Two typical runs, though limited to 100 grains for each $a$ value and with integration times of only 0.1 Myr for clarity, are illustrated in Figs. \ref{self_fig3a} and \ref{self_fig4}. The spread of the orbits into broom-like structures is entirely due to the turbulence; if ${\bf v}_{\rm turb}$ is set 
equal to ${\bf 0}$ the orbits collapse to a single vertical line. It is obvious from these figures that the choice of initial conditions for the velocity is important for the orbits, although the fraction of grains falling into the star as a function of grain radius is not as dependent on these conditions. 

The main results of these numerical experiments are displayed in Fig. \ref{self_fig5}. Here we have plotted a quantity called  `the defect',} $ D(a),$
of the run, as a function of $a$ for different choices of the defining parameters. $D(a)$ measures the fraction of grains that leave the proto-solar 
cloud as a result of the radiative cleansing. It is defined as
\begin{equation}
D(a) = 1 - \frac{N_{\rm l}}{N_0},
\label{eq70}
\end{equation} 
where $N_{\rm l}$ is the number of dust grains lost inside 100 AU, and $N_0$ is the corresponding number lost in a run with identical parameters with the
exception that the stellar luminosity $L$ is set equal to zero. In this way, we try to correct for the fact that even at zero luminosity some of the gas may depart at high
turbulence speeds from the stellar neighbourhood. It is seen from the figure that the variation of $D$ with $a$ is smoothed when $v_{\rm turb}$ increases, from  a very steep change when $a$ passes the limiting value in Equation (\ref{eq2}) for small turbulence, to a less steep variation. We also note that an increased 
inflow of gas decreases the deficit at a given $a$, as does a change of the initial condition from ${\bf v(r_0)}={\bf 0}$ to setting ${\bf v(r_0)}$ equal to
the free fall velocity of the gas flow. Also, the coulomb forces diminish $D$ but not drastically so. None of these effects is unexpected. Obviously, these variations of the deficit with the different parameters are not very dramatic: the conclusions concerning the cleansing effects in our model are clearly relatively robust with regard to its uncertain parameters. This may be seen by calculating  the integrated deficit,} $\mathcal{D,}$ as
\begin{equation}
\begin{split}
\mathcal{D} \equiv \left. \int_{a_l}^{a_u} D(a) \frac{4 \pi}{3} a^3 \rho_d N_0 a^{-3.5} da \middle/ \int_{a_l}^{a_u} \frac{4 \pi}{3} a^3 \rho_d N_0 a^{-3.5} da \right. \\
 =  \left. \int_{a_l}^{a_u} D(a) a^{-0.5} da \middle/ \int_{a_l}^{a_u} a^{-0.5} da \right..
\label{eq91}
\end{split}
\end{equation}
$\mathcal{D}$ measures the fraction of the dust mass that is pushed away from the star instead of falling into it. In Table \ref{tab1} we present 
the values of $\mathcal{D}$ for our different models in Figure \ref{self_fig5}, and some other models. The inclusion of electrical forces and in particular the photo-electric charging obviously
leads to a moderate decrease of $\mathcal{D}$. Also, 
turbulence-generated density variations lead to
a not very dramatic shift of  $D(a)$ to smaller values (model no. 10 in the table). As seen in the table (model no. 11), an adopted variation of the density as $1/r^2$, still assuming 
the inflow gas velocity to be roughly proportional to $1/r^{3/2}$  which means that the gas is piling up in the inner region of the
gas envelope, leads only to small effects compared with the standard stationary flow models with $\rho \propto r^{-3/2}$. It seems that a rather robust conclusion may be drawn: about half the mass
of dust in a late accretion flow towards a solar-type star is cleansed from the flow by the radiative forces of the star itself, as long as the in-falling gas mass rate
is not higher than about 0.01 M$_\odot$/Myr.  

\begin{figure}
  \resizebox{\hsize}{!}{\includegraphics{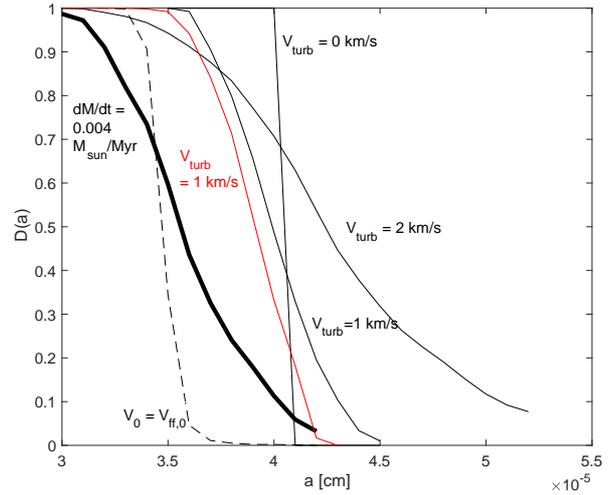}}
  \caption{The defect, $D(a),$ (cf. Equation (\ref{eq70}) as a function of grain radius $a$ for different values of the turbulence speeds $v_{\rm turb}$ and with initial dust velocities = ${\bf v}_0 = {\bf 0}$ and with infall rate of gas ${\dot M}=0.002$ M$_{\odot}$/Myr shown in thin lines. Red line: Coulomb forces, excluding photo-electric effect, taken into account with an ionization ratio 
  of $n(e)/n(H) = 2.72\cdot 10^{-4}$.Thick line: 
doubled infall rate, $v_{\rm turb}=1$ km/s. Dashed line: initial speed ${\bf v}_0 =$ initial free-fall velocity of the gas, ${\bf v}_{\rm turb} = 1$ km/s, ${\dot M}=0.002$ M$_{\odot}$/Myr.}
 \label{self_fig5}
\end{figure} 
  
\begin{table}
\caption{Values of the integrated deficit, $\mathcal{D,}$ for runs of models with turbulence. 
The parameters of the models are the 1D turbulence parameter $v_{\rm turb}$ in km/s, and the gas-flow rate $\dot{M}$ in M$_\odot$/Myr. The initial condition for
the dust grains at distance $r_0=$ 10,000 AU from the star is ${\bf v}_0$ = ${\bf 0}$ except for a model where ${\bf v}_0$ is set to the free-fall velocity ${\bf v}_{\rm ff}$ of
the gas at $r_0$. All models were run neglecting the coulomb forces, except two, marked `coulomb' and `+ photo-el'. For these, the coulomb forces, 
without and with the photoelectric charging of the dust grains, respectively, were included schematically (see text). 
The model marked $\delta\rho$ in the last column was run with fluctuating density, see text, and the following model with a $\rho\propto r^{-2}$
dependence and $r_0=1000$ AU. 
The two last lines represent runs where the star was enclosed in a 0.125 pc$^3$ cubic box with six reflecting walls, to simulate the contributions of new grains
from surrounding stars, and five reflecting walls, respectively.
}        
\label{tab1}      
\begin{tabular}{c c c c c l}     
\hline\hline       
Model &${\bf v}_{\rm turb} $ & $\dot{M}$ & ${\bf v}_0 $& $\mathcal{D}$ & note \\
 & km/s & M$_\odot$/Myr &&\\
\hline \\
$1$&$0$&$0.002$&${\bf 0}$&$0.61$\\              
$2$&$1$&0.002&${\bf 0}$&$0.61$\\
$3$&$1$&0.002&${\bf 0}$&$0.60$& coulomb\\
$4$&$1$&0.002&${\bf 0}$&$0.47$&   + photo-el.\\
$5$&$2$&0.002&${\bf 0}$&$0.64$\\
\hline
$6$&$1$&$0.004$&${\bf 0}$&$0.55$\\
$7$&$1$&$0.010$&${\bf 0}$&$0.41$\\ 
$8$&$1$&0.002&${\bf v}_{\rm ff}$&$0.54$\\ 
$9$&$1$&0.002&${\bf 0}$&$0.60$& $r_o = 1000$AU\\
$10$&$1$&0.002&${\bf 0}$&$0.58$& $\delta \rho$\\ 
$11$&$1$&0.002&${\bf 0}$&$0.60$& $\rho^{-2}, r_o = 1000$AU\\
\hline
$12$&$1$&0.002&${\bf 0}$&$0.60$ &closed box\\ 
$13$&$1$&0.002&${\bf 0}$&$0.61$ & semi-closed box\\ 
\hline 
\end{tabular}
\end{table}
The present model is unrealistic in several respects. One of the most important simplifications
relative to reality is the
neglect of magnetic fields. The star-forming clouds and the protostars are known to be magnetic (see e.g. \citet{Higuchi18}, \citet{Hull18} and references therein), which may have consequences for the
dynamics of the in-falling gas. Also, since the grains may be charged, their dynamics
may be directly affected by the fields. We have addressed the question of whether the fields threading the gas are swept along by it by estimating the timescale for ambipolar diffusion of the field out of the gas,
following Equation (2.18) of  \citet{Hennebelle12a}, which shows that the diffusion time  varies as $n\cdot n_i \, (L/B)^2$ 
where $n$ and $n_i$ are the number densities of atoms and of ions, respectively, 
$L$ is a characteristic length scale and $B$ is the field strength. For our extended and thin flows ($L \sim 1000 \, {\rm AU}, n \sim 0.3 \, 10^3 {\rm cm}^{-3}$  and for characteristic numbers
of the ionization fractions in the cold gas and 
with $B$ set to a few microGauss (\citet{Padoan18}), diffusion timescales are at least one order of magnitude 
greater than $10^6$ years. 
Therefore, the magnetic fields are swept along by the in-falling gas. Our estimates of the magnetic-field energy 
density $E_B$, as compared with the kinetic 
energy density $E_K$ of the gas and its gravitational energy density $E_G$, show that $E_B << E_K \sim E_G$,
which suggests that the magnetic effects on the gas dynamics are only marginal.  As regards the direct effects of 
the magnetic fields on the dynamics of the grains, they may, however be very considerable. According to 
\citet{Draine11} the dynamics is determined by the dimensionless quantity $\omega \cdot \tau_{\rm drag}$ where $\omega$ is
the gyro-frequency (proportional to the field strength and the charge-to-mass ratio of the grain), and $\tau_{\rm drag}= (m\cdot v)/F_{\rm drag}$ is the drag timescale for the grain with a drag force $F_{\rm drag}$, a mass $m$ and a speed $v$. 
From Equation (30) of \citet{Draine11}, we find with our characteristic numbers that $\omega \tau_{\rm drag}$ may be 
typically in the range 1 - 10, which means that the magnetic fields, if oriented at angles relative to the
forces on the grain, may significantly affect its motions. In particular, if the magnetic field is perpendicular
to the direction from the grain towards the star, with the  radiative force on the grain dominating relative to gravity
and the gas drag force,  the grain will nevertheless be dragged 
along in the inflow motion of the gas by the magnetic field.
Key factors affecting the degree to which magnetic fields may thus hamper the radiative dust cleansing are therefore (1) the number of magnetic field lines in the outer regions of the gas (at $r \sim 10^3$ AU) in late 
and thin accretion flows that are close to perpendicular to the flow lines and (2) how stationary such 
regions are, that are locking the grains to the collapsing gas; that is, if 
the characteristic time for reorientation of the field in the gas due to turbulence is smaller or greater than the 
free-fall time of the gas towards the star. The study of such questions requires more detailed numerical MHD 
simulations.

The effects of angular momentum are also neglected in the present simple model, which may be relevant since much of
the dust-gas separation takes place at relatively great distances from the central star. However, closer to the star the 
orbits of the grains may be quite different due to their intrinsic angular momentum, and the non-radial drag of the rotating gas with 
the magnetic field. If the
magnetic field drags the dust inwards from large distances, the winding-up of the field lines in the accretion disk may finally
decelerate dust as it falls  inwards, and outflows at angles with the disk field may even accelerate the grains selectively outwards again. 

For dust located in the inner part of the disk, the radiative field from the star
may be very severely reduced by extinction. In this case, the pressure by infrared radiation from the hot disk may 
dominate and cause large dust grains to slide above the surface of the disk to larger radii (\citet{Vinkovic09}) while smaller grains are less affected. As 
suggested by the referee of the present paper, for grains orbiting the 
star, the Poynting-Robertson drag will tend to selectively decelerate the smaller grains and make them move 
inward towards the star. Both these effects may therefore favour larger grains at greater stellar distances as compared
with smaller grains, which is contrary to the mechanism discussed in the present paper. As highlighted above, the composition and structure of the grains
are also expected to affect the $q$ value in Equation (\ref{eq1}); flows with small graphite grains and relatively large fluffy composites may for that reason also be 
expected to be more effectively radiatively cleansed.

\subsection{Stellar fields}

Above, we have studied the case of a single star with a relatively mild flow of gas falling in towards it. The integrations of the dust orbits started
at considerable distance (10,000 AU) from the star, and the gas flow was assumed to continue in a stationary way for several million years, which implies
that regions far outside this distance from the star may also be involved. In reality,
such an isolated situation should be rare: stars are known to be formed in associations or clusters, and the effects of the companion stars should be considered. 
In the case of the Sun, it is often argued that it had its origin in a relatively dense stellar environment on the basis of the abundances of some daughter elements of 
radioactive supernova products like $^{26}$Al, $^{41}$Ca and $^{60}$Fe. The situation has been summarised by \citet{Adams10} who also included
various dynamic arguments in his discussion and concluded that the number of members of the solar birth cluster should fall in the range $N\sim 10^3 - 10^4$, which
suggests a cluster radius of about 6 pc (cf. \citet{Joshi16}), and a typical distance to the closest star of about 0.5 pc. We note, however, the recent discussion by 
\citet{Fujimoto18} who argue that "normal" chemo-galactic evolution with new generations of stars preferentially forming in patches of the Galaxy that were
contaminated by previous generations of stars can account for the observed radioisotopes without invoking a nearby supernova very close to the forming Solar system. 
Nearby stars and star passages may episodically affect the dynamics of the gas flows towards the accreting star. However, such effects on $\mathcal{D}$ are probably not very
significant, in view of the relatively small effects of the persistent turbulence found above. However, if the neighbouring stars are assumed to actively cleans their 
envelopes, similarly to and simultaneously with the corresponding self-cleansing of the Sun, we expect flows of radiatively driven grains from those stars towards the 
Sun. Therefore, a reduction of $\mathcal{D}$ for the Sun is possible. 

We explored the effects of the dust from stellar neighbours by surrounding the star in the centre by a box  of $d \times d \times d$ pc$^3$ in dimension, and then 
reflected the velocities of the outgoing particles when they hit the walls of the box. In order to avoid biases, we then shifted the starting point of the grain
on the box side randomly (a precaution which was found to be of little significance for the result). The reflection was repeated if the grain hit another 
box surface until the full integration time of 5 Myr had elapsed. As is seen in Table \ref{tab1}, the effects of this closed box around the star are relatively small: the 
value of $\mathcal{D}$ is reduced by only one percent, and, in fact, does not significantly depend  on the value of $d$. The reason for the small magnitude of these effects is the 
symmetry in the forces dominating at a given distance from the star, radiative pressure and gravity, which both scale as $1/r^2$, for a dust grain that departs from, or 
approaches, the star. Only when the grain is much closer (with the mild accretion rates adopted here), does friction introduce an asymmetry, so that the speed of the
grain at a given distance from the star becomes different and  higher in the approach than in the departure. For radiative cleansing to be significant at all, the accretion rates should
not be too high, as is illustrated in Fig. \ref{self_fig1}, and this requirement prevents the asymmetries from becoming larger. 

The conclusion of these schematic experiments is that the effects on dust in a stellar envelope of the emission of dust from solar-type stars
in a dense stellar environment should not
be very significant in reducing the cleansing net-effects from those of isolated stars.  One should note, however, that the 
radiation from neighbouring stars may be of considerable significance, not by pushing dust grains from their own envelopes towards the Sun but by
helping in pushing the dust away from the accreting Sun and thus increasing $\mathcal{D}$. Nevertheless, effects
of the gas flows from stellar winds and expanding H\,II regions may be equally significant, and in practice reduce the final accretion of dust-poor 
matter onto the Sun. Obviously, this complex situation requires more detailed modelling, and the results may well turn out to be quite dependent on initial conditions and
the geometrical characteristics of the arrangement. 

\subsection{The inflow: its rate and anisotropy, and the accretion disk}
In order for the proposed mechanism to lead to the consequences observed, that is, a depletion of the dust-forming elements in the 
solar atmosphere of $\sim 20\%$, the inflow rate of the gas (assumed here to correspond to the accretion rate via the
accretion disk) multiplied by the time $t_{\rm flow}$when the flow is active must match the mass $m_{\rm conv}$ of the stellar convection zone such that
\begin{equation}
\dot{M} \times t_{\rm flow}  > 0.2 m_{\rm conv}. 
\label{eq1010}
\end{equation}
Adopting the mass of the convection zone following Figure 4 of \citet{Baraffe10} for a non-accreting 1 M$_\odot$ model, we find typical
convective masses of 0.5, 0.2, and 0.05 M$_\odot$ at 10, 20, and 30 Myr, respectively. With those numbers, the condition of Equation (\ref{eq1010}) 
would be fulfilled, even with significantly smaller flow rates $\dot{M}$ than 0.02 M$_\odot$/Myr, provided that these remain relatively 
constant during the respective times. 

Star-forming gas is far from homogeneous. Instead, these gas clouds show cores, sheets, and filaments. This is clearly displayed in the simulations
by \citet{Kuffmeier17} which demonstrate that stars accrete gas often streaming in filaments that simply fill a part (typically about 10 \%)\ of the full
space angle as seen from the central region with the protostar and its accretion disk (see their Figure 7).  If such inhomogeneities are also characteristic 
of the much later accretion stages hypothesised here, the inflow rates in the filaments must be increased correspondingly, relative to those of the 
isotropic case discussed above. Such an increase is still possible within the conditions set by Equation ({\ref{eq33}) and would admit a small optical
depth in the filament. 

The basic conclusion of the present study is that the early Sun may have radiatively cleansed its outer layers from refractory elements if the accretion rates
in late phases were small but significant. 
An interesting question would then be whether or not an accretion disk should be observable around stars also at these stages, far beyond
the generally quoted observed lifetimes of protoplanetary disks of, at the most, 10 Myr (see \citet{Mamajek09},  \citet{Williams11} and references therein, but also \citet{Pfalzner14} who point out
that the lifetimes may be underestimated). \citet{Kraus14}, on the basis of IR observations, find an upper limit of 5\% on the fraction of K3.0-K7.9 stars in the Tuc-Hor moving group (with an estimated age of 30-45 Myr) to host
significant amounts of warm circumstellar dust. According to \citet{Alexander14} "a small (but not negligible $\lessapprox 10\%$) sample of older $ \sim10-20$ Myr pre-main
sequence stars ... retain their dust (and gas) signatures". 

The possible existence of disks at small accretion flows is dependent on the outflow rates from the disk region and the replenishment time for the disk. 
The outflow, whether due to photo-evaporation by EUV, FUV,  X-rays, or MHD disk winds (see \citet{Alexander14} for a discussion on this controversial issue)
may be varying from star to star, but is believed to finally disperse the disk and end the accretion.  If the renewal time is also short  for late
stages, for example $\sim 
30,000$ yr as found in the simulations by \citet{Kuffmeier17}, the global accretion rates of our model suggest disk masses of approximately $10^{-4}$ M$_\odot$ at most,
while with the longer dissipation times, measured from the time when accretion stops as estimated by several authors (see \citet{Williams11} for references)
we could find disks that are more massive by  one order of magnitude. 
\citet{Hartmann16} (their Figure 8) plotted observed accretion rates based on spectroscopic observations of the Balmer continuum and emission lines versus
estimated age.  We note that for stars with ages between 10 and 30 Myr the rates are of the same order of magnitude as those needed for our mechanism to work, although the data have great scatter and probably suffer from severe bias. It may be rewarding to look further for accretion flows and disks around solar-type stars in associations with ages of 20-30 Myr.


\section{Conclusions}

In spite of the exploratory nature of our study of the possibility of `self-cleansing' of the dusty envelope around the early Sun, 
we can safely conclude that the proposed mechanism would cleanse the gas of about half its dust mass early on,
provided that the incoming gas flow is not much more massive than  0.01 M$_\odot$/Myr, and is sufficiently homogeneous, non-magnetic, and non-rotating.
In a dense cluster the situation is uncertain, however, and probably dependent on the geometry of 
the stellar field, the presence of luminous stars, and of outflows and winds, and so on. Also, the presence of outflows from the early Sun itself
may affect the situation considerably. Moreover, we have neglected the inner regions of the solar nebula, where the proto-planetary disk 
certainly might complicate the situation, both dynamically, via the gas flow and the magnetic field and by shielding the outer envelope from stellar radiation in certain directions, as well as by 
promoting other dust-fractionation processes. In order to clarify the effects of such complications, detailed MHD simulations, partly analogous to those of more massive stars
by \citet{Wibking18} and those allowing for angular momentum,  should be performed.

One may question whether the mechanism proposed could be used to explain the findings by several research groups mentioned in the Introduction, that is, of binary stars with 
significantly different compositions, and cases with significant trends in these differences with condensation temperature
of the elements. A natural way of obtaining the observed differences within the framework of our cleansing paradigm would be to invoke
a situation in which the two components of the binary had different accretion rates, one with Equation (\ref{eq33}) satisfied and one where the rate was
higher, for example, as a result of an inhomogeneous distribution
of the inflowing gas. An alternative explanation might be 
that the two components were in somewhat different evolutionary phases, with convection zones of significantly different depth, when this 
final accretion occurred, meaning that the
accreted dust-cleansed material was mixed with different amounts of stellar material. 

The finding by \citet{Gonzalez10} that CI meteorites show even lower refractory/volatile abundance ratios than the solar photosphere would also be
explained by the present mechanism if the radiatively cleansed matter from which the meteorites formed was not enough to replace all the convection
zone but was mixed into it.  

The seemingly systematic low abundance ratio of
refractory elements to volatile elements of the cluster stars in M\,67 seems, however, more difficult to relate to the present scenario. 
One hypothesis might be that the deep gravity well of the cluster retained gas for a long time, which could then slowly accrete onto the stars with self-cleansing
in operation. Another possibility is that the radiative pressure from the bright stars in the cluster could drive away the dust grains from the late accretion flows of the less massive stars. In both these
scenarios one would expect the ratio of refractories to volatiles in the stellar atmospheres to increase towards the lower end
of the main sequence of the cluster, as a result of the deepening of the convection zone with decreasing temperature.

\begin{acknowledgements}
Martin Asplund, Alexis Brandeker, Jorges Mel{\'e}ndez, Dhruba Mitra, Poul Erik Nissen, {\AA}ke Nordlund, Henk Spruit and an anonymous referee
are thanked for many valuable suggestions. 
\end{acknowledgements}

\bibliographystyle{aa} 
\bibliography{Gustafsson2017.bib}
\end{document}